\documentclass[twocolumn,prl,amsmath,amssymb,floatfix,reprint,square,comma,numbers]{revtex4-1}
\usepackage{graphicx}
\usepackage{dcolumn}
\usepackage{bm}
\usepackage{braket}
\usepackage{amsmath}
\usepackage{textcomp}
\usepackage{color}

\usepackage[colorlinks,citecolor=blue,linkcolor=blue]{hyperref}

\begin{document}

\title{Simultaneous excitation of two noninteracting atoms with time-frequency correlated photon pairs in a superconducting circuit}

\author{Wenhui Ren$^{1,*}$, Wuxin Liu$^{1,*}$, Chao Song$^1$, Hekang Li$^{1,2}$, Qiujiang Guo$^1$, Zhen Wang$^1$, Dongning~Zheng$^{2,3,\dagger}$, Girish S. Agarwal$^{4}$, Marlan O. Scully$^{4,5}$, Shi-Yao Zhu$^1$, H. Wang$^{1,\ddagger}$, and Da-Wei Wang$^{1,3,\S}$}

\affiliation{$^1$Interdisciplinary Center for Quantum Information,~State Key Laboratory of Modern Optical Instrumentation,~and Zhejiang Province Key Laboratory of Quantum Technology and Device,~Department of Physics,~Zhejiang University,~Hangzhou 310027,~China\\
$^2$Institute of Physics, Chinese Academy of Sciences, Bejing 100190, China\\
$^3$CAS Center for Excellence in Topological Quantum Computation, University of Chinese Academy of Sciences, Bejing 100190, China\\
$^4$Institute for Quantum Science and Engineering, Departments of Biological and Agricultural Engineering, Physics and Astronomy, Texas A$\&$M University, College Station 77843, Texas, USA\\
$^5$Baylor University, Waco, Texas 76706, USA}

\begin{abstract}
 Here we report the first observation of simultaneous excitation of two noninteracting atoms by a pair of time-frequency correlated photons in a superconducting circuit. The strong coupling regime of this process enables the synthesis of a three-body interaction Hamiltonian, which allows the generation of the tripartite Greenberger-Horne-Zeilinger state in a single step with a fidelity as high as 0.95. We further demonstrate the quantum Zeno effect of inhibiting the simultaneous two-atom excitation by continuously measuring whether the first photon is emitted. This work provides a new route in synthesizing many-body interaction Hamiltonian and coherent control of entanglement.
\end{abstract}

\maketitle

Two-photon absorption, where two photons successively induce a quantum transition, was predicted by Goeppert-Mayer in 1931 \cite{Mayer1931}. It was not experimentally observed \cite{Kaiser1961} until the invention of the laser, which provides the necessary intense monochromatic radiation. This effect has been widely used in fluorescent imaging \cite{Denk1990, So2000}, microfabrication \cite{Maruo1997}, and quantum light sources \cite{He2019a}. It has been found that specially correlated two photons can excite otherwise impossible two-photon transitions \cite{Muthukrishnan2004}.
In addition to the two-photon absorption in a single atom, it has been found that two photons can also simultaneously excite two atoms that are interacting with each other \cite{White1981,Varada1992,Penafiel2012}, which offers new methods in collective control of quantum systems \cite{Anno2006} and in spectroscopy with nanoscale resolution \cite{Hettich2002}. However, the requisite of interaction between atoms hinders the application of this effect in quantum information processing, since it is difficult to make far apart qubits interact with each other \cite{Bienfait2019,Samutpraphoot2020}. To circumvent this obstacle, many proposals \cite{Garziano2016,Zhao2017,Stassi2017,Liu2019} on simultaneous two-atom excitation have emerged within the framework of cavity quantum electrodynamics. In particular, it has been shown that a single photon can simultaneously excite two atoms \cite{Garziano2016} with the effect of the counter-rotating terms and the ultrastrong coupling \cite{Kockum2019}. However, it is still challenging to realize the ultrastrong coupling and to mitigate its side effect. Until now, the simultaneous excitation of two noninteracting atoms has never been observed experimentally.

While two uncorrelated photons with frequencies $\nu_1$ and $\nu_2$ cannot simultaneously excite two noninteracting atoms with frequencies $\omega_1$ and $\omega_2$ at the two-photon resonance,
\begin{equation}
\begin{aligned}
\nu_1+&\nu_2 =\omega_1+\omega_2,\\
\nu_i\ \neq\omega_j &\text{ for all } i,j=1,2,
\end{aligned}
\label{tpr}
\end{equation} 
it has been shown that the photon pairs generated in a cascade three-level system can excite two noninteracting atoms by breaking the symmetry in the arrival time of the two photons \cite{Muthukrishnan2004, Zheng2013}. However, the difficulty in collecting the photon pairs in free space greatly limits the efficiency of this process despite their promising applications in quantum spectroscopy \cite{Schlawin2013,Dorfman2016}. In this letter, we experimentally demonstrate that by confining the photon pairs in a cavity, two noninteracting artificial atoms can be simultaneously excited in a superconducting circuit. The three-body interaction between the light source and the two atoms is in the strong-coupling regime, such that the Greenberger-Horne-Zeilinger (GHZ) state \cite{Greenberger1990}  can be dynamically generated in a single step. We also show that this many-body dynamics can be coherently controlled by the quantum Zeno effect \cite{Mirsra1977,Itano1990,Fischer2001}.
This scheme requires neither counter-rotating terms nor ultrastrong coupling and provides a new method of synthesizing three-body interactions and entanglement generation. 

\begin{figure}[t]
	\centering
	\includegraphics[width=1\linewidth]{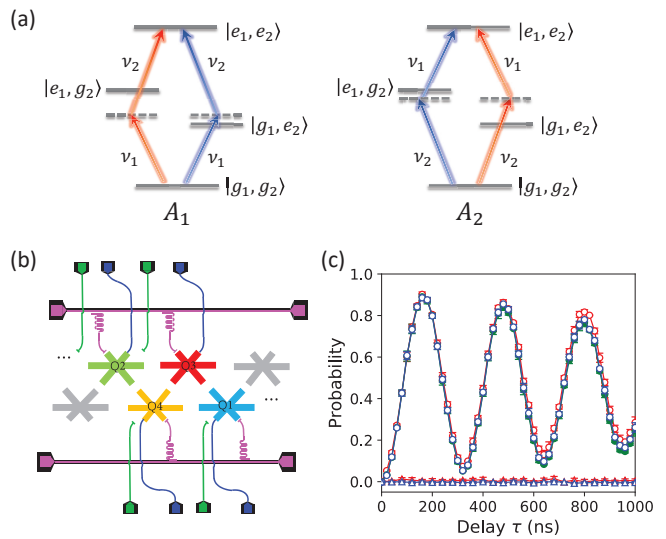}
	\caption{Simultaneous excitation of two noninteracting qubits. (a) The quantum paths for $A_1$ and $A_2$ in Eq. (\ref{tpr}). The pathways with the same color (blue or red) in $A_1$ and $A_2$ cancel each other, evident from their opposite values of the detunings from the atomic levels.
(b) Sketch of part of the multiqubit superconducting circuit. Four qubits, labelled $Q_{1}$, $Q_{2}$, $Q_{3}$ and $Q_{4}$, are used in this experiment. 
%$Q_1$ and $Q_2$ are the two noninteracting qubits. $Q_3$ provides the correlated photon pairs in a cascade transition. $Q_4$ is used to measure the state of $Q_3$ in the quantum Zeno effect. 
Each qubit has its own flux bias line (blue) for fast frequency tuning and microwave line (green) for $SU(2)$ spin rotations. 
For qubit readout, each qubit is capacitively coupled to its own readout resonator which is coupled to one of the two common transmission lines (pink).
(c) The single-qubit probabilities of finding $Q_1$ (blue) and $Q_2$ (red) in their excited states when they are simultaneously driven by correlated photon pairs (circles) and uncorrelated classical microwaves (triangles) as functions of the interaction time $\tau$. {The joint probability of finding $Q_1$ and $Q_2$ in $|e_1,e_2\rangle$ by absorbing the correlated photon pairs is denoted by the filled green circles, which almost overlap with the empty blue and red circles. The simultaneous excitation of the two qubits is in conformity with the prediction of Ref. \cite{Muthukrishnan2004}.}
\label{scheme}
}
\end{figure}

We first illuminate the underlying physics of simultaneous excitation of two noninteracting atoms with time-frequency correlated photon pairs. Intuitively, it is easy to understand that two uncorrelated photons cannot simultaneously excite two noninteracting atoms under the condition in Eq. (\ref{tpr}). Since each atom independently interacts with the two photons and can be considered separately, they cannot be excited because neither of them is resonant with the photons. For the convenience of the discussion on the correlated photon pairs, we alternatively regard the two atoms as an entity and the transition probability can be calculated by considering which photon being absorbed first. The two-photon transition matrix element consists of two parts, $A_1$ and $A_2$ corresponding to the quantum pathways where $\nu_1$ and $\nu_2$ are first absorbed, as shown in Fig.~\ref{scheme} (a). From the second order perturbation theory, we obtain \cite{Muthukrishnan2004},
\begin{equation}
\begin{aligned}
A_1=\frac{\kappa_{22}\kappa_{11}}{\nu_1-\omega_1}+\frac{\kappa_{12}\kappa_{21}}{\nu_1-\omega_2},\\
A_2=\frac{\kappa_{21}\kappa_{12}}{\nu_2-\omega_1}+\frac{\kappa_{11}\kappa_{22}}{\nu_2-\omega_2},
\label{me}
\end{aligned}
\end{equation}
where $\kappa_{ij}$ (much smaller than the one-photon detunings $|\nu_j-\omega_i|$) is the coupling strength between the atom and photon with frequencies $\omega_i$ and $\nu_j$. The two-photon resonance in Eq. (\ref{tpr}) renders that $A_1=-A_2$ and the total two-photon transition matrix element is zero.
However, if the two photons are generated from a cascade transition of a three-level atom such that the photon with frequency $\nu_1$ is guaranteed to arrive at the atoms first, only $A_1$ is present and it is generally nonzero \cite{Muthukrishnan2004}. Consequently, the two atoms can be simultaneously excited. However, although time-frequency correlated photon pairs can be generated in nonlinear crystals \cite{Zhao2014} and atomic media \cite{Liu2012}, the simultaneous excitation of two noninteracting atoms has never been experimentally observed due to the difficulty both in finding the proper atoms and in collecting the entangled photon pairs. 

Superconducting circuits offer a versatile platform to verify this interesting process, thanks to the flexibility of the artificial atoms (superconducting qubits) and the strong coupling of the atom-photon interactions. The experiment is performed in a superconducting circuit consisting of multiple qubits interconnected in a zigzag triangular ladder \cite{Liu2020} (see Fig.~\ref{scheme} (b)), where the couplings between neighboring qubits are measured to be $10\sim13\ \mathrm{MHz}$ and otherwise the couplings are weaker by about an order of magnitude.
Each qubit is a frequency tunable transmon circuit, whose sinusoidal potential well hosts multiple energy levels. The lowest three energy levels of $Q_j$ are referred to as the ground state $|g_j\rangle$, the first excited state $|e_j\rangle$, and the second excited state $|f_j\rangle$, the latter of which is only involved in the experiment for $Q_3$. In our experimental setup,
nonadjacent $Q_1$ and $Q_2$ are the two noninteracting atoms, and $Q_3$ is a qutrit that generates the two cascade photons. $Q_4$ is used as a measuring qubit in demonstrating the quantum Zeno effect.  We can tune the resonance frequencies, perform $SU(2)$ rotations and measure the quantum states of these qubits through their own flux bias lines, $XY$ lines and readout resonators, respectively. 
% which are coupled to a transmission line for dispersive measurement. The performance of each qubit at relevant frequency is characterized by 
%We read out the multiqubit state by applying a multi-tone microwave pulse with a length of about 1000 ns, with each tone targeting one of the readout resonators. 
At the experimental frequencies, the coherence performance of these qubits is characterized by the lifetime and Ramsey interference measurements, which yield the typical relaxation time $T_{1}>20\,\mu$s and the Gaussian dephasing time $T_{2}^{*}\approx 1\,\mu$s (see Supplementary Material for detailed information of the device~\cite{supp}). 

\begin{figure}[t]
	\centering
	\includegraphics[width=0.75\linewidth]{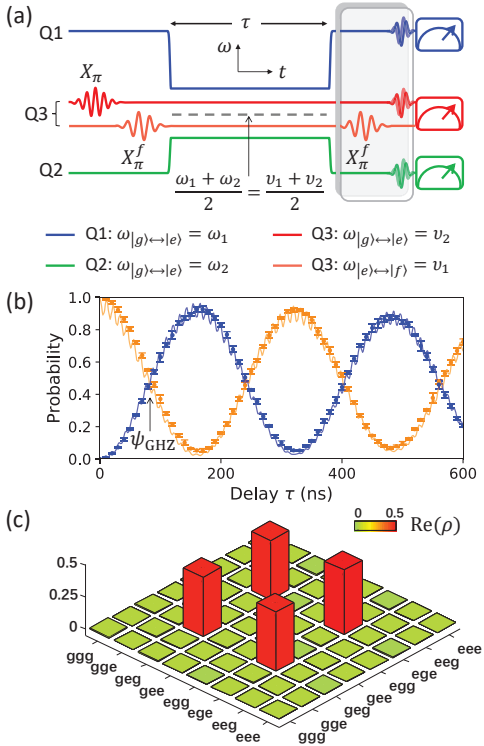}
	\caption{Generating the GHZ state with the three-body interaction Hamiltonian. (a) Illustrative pulse sequence drawn in the frequency versus time frame. After initializing $Q_1$-$Q_2$-$Q_3$ in $|g_1,g_2,g_3\rangle$ at their respective idle frequencies, we prepare the state $|g_1,g_2,f_3\rangle$ by successively applying to $Q_3$ an $X_\pi$ rotation (a $\pi$ rotation around $X$ axis transferring $|g_3\rangle$ to $|e_3\rangle$, red sinusoid) and an $X_\pi^f$ rotation (transferring $|e_3\rangle$ to $|f_3\rangle$, light red sinusoid), following which we apply square pulses to tune the quibt frequencies, so that Eq. (\ref{tpr}) is satisfied, for a dynamical evolution with a variable time $\tau$. Finally we bring the qubits back to their idle frequencies for simultaneous readout: The same pulse sequence is repeated 1500 times to count the 12 probabilities $\{P_{ggg}, P_{gge}, P_{ggf}, \cdots, P_{gef}, P_{eef}\}$. The tomographic pulses in the shaded box is only inserted for QST of the GHZ state, $|\psi_\textrm{GHZ}\rangle$, where the $X_\pi^f$ rotation transforms $|\psi_\textrm{GHZ}\rangle$ to its equivalent $|\psi_\textrm{GHZ}^\prime\rangle=(\left| g_{1},g_{2},e_3 \right \rangle + e^{i\phi} \left| e_{1},e_{2},g_3 \right \rangle)/\sqrt{2}$, followed by null or $\pi/2$ rotations running through the Pauli set $\{I, X_{\pi/2}, Y_{\pi/2}\}$ for all three qubits. (b) The dominant occupational probabilities $|c_1(\tau)|^2$ for $\left| g_1,g_2,f_3 \right\rangle$ ($P_{ggf}$, orange circles) and $|c_2(\tau)|^2$ for $\left| e_1,e_2,g_3 \right\rangle$ ($P_{eeg}$, blue circles) measured as functions of the interaction time $\tau$, in comparison with the numerical simulation (lines). (c) QST of the experimental $|\psi_\textrm{GHZ}^\prime\rangle$ at $\tau \approx 88$~ns. Shown is the real part of the three-qubit density matrix, $\rho$, after a numerical rotation to remove the phase $\phi$ in $|\psi_\textrm{GHZ}^\prime\rangle$.}
\label{GHZ}
\end{figure}

\begin{figure}[t]
	\centering
	\includegraphics[width=0.8\linewidth]{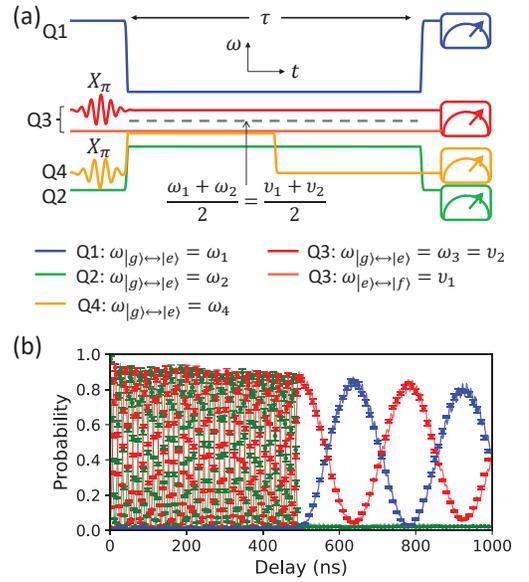}
	\caption{The quantum Zeno effect of inhibiting the simultaneous two-atom excitation by continuous observation. (a) Illustrative pulse sequence drawn in the frequency versus time frame. Starting from all four qubits in their ground state, we first apply two $X_\pi$ rotations to $Q_{3}$ and $Q_{4}$ (red and orange sinusoids). Then, we tune the frequencies of the four qubits into two-photon resonance, i.e., $\omega_1 + \omega_2 = \omega_3 + \omega_4$ where $\omega_3=\nu_2$ and $\omega_4=\nu_1$, by applying square pulses to the qubit flux bias lines. At $\tau < 500$ ns, $Q_4$ is continuously measuring the state of $Q_3$. At $\tau\approx 500$~ns when $Q_3$ is in state $|f_3\rangle$, we tune $Q_4$ to its idle frequency by applying a rectangle pulse. (b) The probabilities of finding the qubits in the state $|g_1,g_2,e_3,e_4 \rangle$ (green circles), $|g_1,g_2,f_3,g_4\rangle$ (red circles) and $|e_1, e_2, g_3, g_4\rangle$ (blue circles) as functions of the interaction time $\tau$, in comparison with the numerical simulation (lines). At $\tau < 500$~ns, we observe the Rabi oscillation between $|g_1,g_2,e_3,e_4\rangle$ and $|g_1,g_2,f_3,g_4\rangle$, without the state $|e_1,e_2,g_3,g_4\rangle$ being noticeably excited due to the constant observation of $Q_3$ by $Q_4$. At $\tau > 500$~ns when $Q_4$ is tuned off-resonant with $Q_3$, we observe the simultaneous excitation of $Q_1$ and $Q_2$ by the two cascade photons from $Q_3$.}
	\label{Zeno}
\end{figure}

The key experimental result is shown in Fig.~\ref{scheme} (c). We first prepare $Q_3$ in the state $|f_3\rangle$, which subsequently emits two cascade photons with frequencies $\nu_1\approx4.865$ GHz (for $|f_3\rangle\rightarrow|e_3\rangle$ transition) and $\nu_2\approx5.100$ GHz (for $|e_3\rangle\rightarrow|g_3\rangle$ transition). These two photons simultaneously excite $Q_1$ and $Q_2$ with transition frequencies $\omega_1\approx5.183$ GHz and $\omega_2\approx4.781$ GHz, which satisfy Eq. (\ref{tpr}). The strong coupling regime of this process is demonstrated by the well observed three-body Rabi oscillation in Fig.~\ref{scheme} (c).
%during which the two excited qubits subsequently excite $Q_3$ in the second half of the period. 
We notice that the probability of finding $Q_1$ and $Q_2$ jointly in $|e_1,e_2\rangle$ (filled green circles) are almost the same as those of finding them individually excited (empty blue and red circles), which is evidence of the simultaneity of the two-atom excitation. In contrast, by simultaneously driving $Q_1$ and $Q_2$ with uncorrelated classical microwave pulses with the same frequencies $\nu_1$ and $\nu_2$, we observe no excitation of the qubits, as shown by the triangles in Fig.~\ref{scheme} (c).

During the process of the simultaneous excitation of the two atoms, three-body entanglement between the light source and the two atoms is generated. This can be seen from the effective Hamiltonian between $Q_1$, $Q_2$ and $Q_3$ ($\hbar=1$),
\begin{eqnarray}\label{He}
H_{\rm eff} = A_1 \sigma_{1}^{+} \sigma_{2}^{+} \sigma_{3}^{-} +h.c.,
\label{3bi}
\end{eqnarray}
where $A_1$ is shown in Eq. (\ref{me}) %with $g_{22}g_{11}=g_{12}g_{21}=\sqrt{2}g_1g_2$ and $g_j$ is the coupling strength between $Q_j$ and $Q_3$. 
and its value is approximately $2$ MHz.
The raising operators for the three qubits are defined as $\sigma_j^+=|e_j\rangle\langle g_j|$ for $j=1,2$ and $\sigma_3^+=|f_3\rangle\langle g_3|$, and the lowering operators are their Hermitian conjugate. The Hamiltonian in Eq. (\ref{He}) is a three-body interaction Hamiltonian that can entangle the three qubits in a single step \cite{Garziano2016,Zhao2017,Stassi2017}, which has never been realized experimentally. In the present scheme, the large value of $A_1$ allows a high-fidelity GHZ state to be generated within a time scale much shorter than the typical coherence times of these qubits (see Supplementary Material~\cite{supp}). The experimental result is shown in Fig.~\ref{GHZ}.

The pulse sequence for generating the GHZ state is shown in Fig.~\ref{GHZ} (a). We first prepare $Q_1$-$Q_2$-$Q_3$ in the initial state $\left| \psi ({0}) \right\rangle = |g_1,g_2,f_3\rangle$ at their respective idle frequencies. Then the three qubits are quickly biased to their operating frequencies, so that Eq. (\ref{tpr}) is satisfied \cite{supp}, by rectangular pulses applied through their flux bias lines. After the dynamic evolution with a variable time $\tau$, we bring the qubits back to their idle frequencies for measurement. 
The results of the joint measurement of the wavefunction $\left| \psi (\tau) \right\rangle =c_1(\tau) |g_1,g_2,f_3\rangle+c_2(\tau) |e_1,e_2,g_3\rangle$, ignoring the insignificant terms, are shown in Fig.~\ref{GHZ} (b), in which the experimentally obtained probabilities of $|c_1(\tau)|^2$ and $|c_2(\tau)|^2$ are plotted as functions of the interaction time $\tau$. At $\tau \approx 88$~ns, ideally a three-qubit GHZ state $|\psi_\textrm{GHZ}\rangle = (\left| g_{1},g_{2},f_3 \right \rangle + e^{i(\pi/2+\phi)} \left| e_{1},e_{2},g_3 \right \rangle)/\sqrt{2}$ is generated, where $\phi$ is the dynamical phase picked up as the three qubits are tuned back to their idle frequencies right after the three-body dynamics. Quantum state tomography (QST) of a $\psi_\textrm{GHZ}$-equivalent state is shown in Fig.~\ref{GHZ} (c), which has a state fidelity of 0.9491$\pm$0.0073 (see Supplementary Materials for details of QST).

Our superconducting circuit allows us to investigate the quantum Zeno effect from the inhibition of the simultaneous excitation of the two qubits by frequently measuring whether the first photon $\nu_1$ is emitted. The quantum Zeno effect was originally proposed as a fundamental question on whether frequent measurements can stop the spontaneous decay of a quantum system \cite{Mirsra1977}. In early experiments, it was demonstrated in quantum systems undergoing coherent evolution \cite{Itano1990} and barrier tunneling \cite{Fischer2001}. It was later found that the quantum Zeno effect is closely related to the dynamical decoupling and can be used to control the quantum coherence and entanglement \cite{Facchi2004,Maniscalco2008}. This has been recently implemented in both natural \cite{Signoles2014,Patil2015} and artificial atoms \cite{Harrington2017,Gourgy2018,Maier2019}. Here we demonstrate that the three-body interaction can be effectively inhibited by frequently measuring its initial state. This measurement is conducted by resonantly coupling qubit $Q_4$ to the transition $|f_3\rangle \leftrightarrow |e_3\rangle$ of $Q_3$, i.e., by setting $\omega_4\approx\nu_1$, so that $Q_3$ in $|f_3\rangle$ can excite $Q_4$ in about 14~ns, which is an order of magnitude shorter than the period of the three-body Rabi oscillation. By continuously observing $Q_3$, the three-body Rabi oscillation is totally inhibited, as shown in Fig.~\ref{Zeno}. In contrast, once the observation is stopped by tuning $\omega_4$ far away from $\nu_1$, the three-body Rabi oscillation revives.

The pulse sequence of the four participating qubits in the quantum Zeno effect is shown in Fig.~\ref{Zeno} (a). Initially, the four qubits are tuned at their respective idle frequencies and are prepared in the state $|g_1,g_2,e_3,e_4\rangle$ by two $X_\pi$ rotations simultaneously applied to $Q_{3}$ and $Q_{4}$. %, which have Gaussian envelops with a full width at half maximum of 30 ns. 
Then the four qubits are quickly biased to their operating frequencies by rectangular pulses applied through their flux bias lines. 
At $\tau < 500$~ns in Fig.~\ref{Zeno} (b), the frequencies of the four qubits are set to satisfy Eq. (\ref{tpr}) and $\omega_4\approx\nu_1$. The four qubits undergo a quick Rabi oscillation between the states $|g_1,g_2,e_3,e_4\rangle$ and $|g_1,g_2,f_3,g_4\rangle$. We observe that although $Q_3$ has a 50\% probability to be found in $|f_3\rangle$, it stops exciting the two qubits $Q_1$ and $Q_2$ due to the continuous observation by $Q_4$. We also notice that the state $|g_1,g_2,e_3,e_4\rangle$ has negligible coupling with $|e_1,e_2,g_3,g_4\rangle$, so that $Q_4$ does not have any direct effect on $Q_1$ and $Q_2$, which is ideal for the quantum Zeno effect. At $\tau > 500$~ns, $Q_1$, $Q_2$ and $Q_3$ remain unchanged, but $Q_4$ is tuned to its idle frequency. As expected, without the measurement from $Q_4$, the two qubits $Q_1$and $Q_2$ are subsequently excited by $Q_3$.

The simultaneous excitation of two noninteracting atoms is of fundamental importance to the basic law of physics. Since the two atoms are not required to have interactions, in principle they can be placed arbitrarily far apart \cite{Muthukrishnan2004}. This indicates that the law of energy conservation can be locally violated (the absence of the one-photon resonance), although the energy is still conserved globally (i.e., the two-photon resonance). This is achieved by breaking the symmetry in the arrival time of the two photons. Since local energy conservation originates from the time-translational symmetry according to the Noether's theorem, the asymmetric arrival time of the two photons relieves the constraint of the local energy conservation. This effect also offers new freedom in engineering many-body interaction Hamiltonians, complementing the two-body interaction Hamiltonians which have been intensively investigated and utilized in quantum simulation \cite{James2007,Xu2018,Wang2019,Song2019}. A Hamiltonian that contains the three-spin chirality operator has been synthesized with Floquet modulation \cite{Liu2020}, which nevertheless complicates the controlling procedure. Here the three-body Hamiltonian in Eq. (\ref{3bi}) does not resort to Floquet modulation, and demonstrates a high fidelity in generating entanglement. Following the same line, four-body and five-body interaction Hamiltonians can also be synthesized and the current multiqubit superconducting circuit can be used to demonstrate novel topological orders \cite{Dai2017,Luo2018}.

We acknowledge the support of the National Science Foundation of China (Grants No. 11934011, No. 11725419, No. 11874322), the National Key Research and Development Program of China (Grants No. 2019YFA0308100, No. 2017YFA0304300), the Zhejiang Province Key
Research and Development Program (Grant No.
2020C01019) and the Basic Research Funding of Zhejiang University. We also gratefully acknowledge the support of AFOSR Award
No. FA-9550-18-1-0141, ONR Award No. N00014-16-1-
3054, and the Robert A. Welch Foundation (Awards No.
A-1261 and No. A-1943-20180324) and the KUAST award. This research is also supported by King Abdulaziz City for Science and Technology (KACST)\\

\noindent
$^*$These authors contributed equally to this work.\\
$^\dagger$dzheng@iphy.ac.cn\\
$^\ddagger$hhwang@zju.edu.cn\\
$^\S$dwwang@zju.edu.cn\\

\end{document}